\documentstyle[sprocl]{article}

\def\be{\begin{equation}}
\def\ee{\end{equation}}
\def\bea{\begin{eqnarray}}
\def\eea{\end{eqnarray}}

\begin{document}

\title{BARYOGENESIS AT THE QCD SCALE\footnote{Based on a plenary talk by R.B. at SEWM-98 and an invited talk by A.Z. at COSMO-98, to be publ. in the proceedings of SEWM-98 (World Scientific, Singapore, 1999)}}

\author{R. Brandenberger$^{1,2)}$, I. Halperin$^{2)}$ and A. Zhitnitsky$^{2)}$}

\address{~\\$^1$Department of Physics, Brown University,\\ 
Providence, RI 02912, USA, E-mail: rhb@het.brown.edu;\\
~\\$^2$Department of Physics and Astronomy, University of British Columbia,
Vancouver, BC, V6T 1Z1, CANADA}


\maketitle\abstracts{ We propose a new mechanism for explaining the observed asymmetry between matter and antimatter, based on nonperturbative physics at the QCD scale. Our mechanism is a charge separation scenario, making use of domain walls separating the recently discovered long-lived 
metastable vacua from the lowest energy vacuum. The 
walls acquire a fractional negative baryon charge, leaving 
behind a compensating positive baryon charge in the bulk. The regions of metastable vacuum bounded by walls (``B-shells") will contribute to the dark matter of the Universe.}

\section{Introduction}

The origin of the asymmetry between baryons and antibaryons, 
and, more specifically, the origin of the observed baryon to
entropy ratio $n_B / s \sim 10^{-10}$ ($n_B$ being the net 
baryon number density, and $s$ the entropy density) remains 
a mystery and one of the main challenges for particle-cosmology. In order to explain this number from symmetric initial conditions in the very early Universe, it is generally assumed that three criteria, first laid down by Sakharov \cite{Sakharov} must be satified:
\begin{itemize}
\item{} Baryon number violating processes exist.
\item{} These processes involve C and CP violation.
\item{} The processes take place out of thermal equilibrium.
\end{itemize}

Here we report on a recent proposal \cite{BHZ} that baryogenesis may be realized at the QCD phase transition. This scenario is based on the existence of domain walls separating the recently discovered \cite{HZ} long-lived metastable vacua of low-energy QCD from the stable vacuum. The walls acquire a negative fractional baryon charge, leaving behind a compensating positive baryon charge in the bulk. In this sense, our proposal is a charge separation rather than a charge generation mechanism.

It is well known that topological defects can play an important role in baryogenesis. This was already realized \cite{BDH} in the context of GUT scale baryogenesis \cite{BG}. In grand unified theoreis (GUT), baryogenesis can occur at or immediately after the symmetry breaking phase transition via the out-of-equilibrium decay of the superheavy Higgs and gauge particles $X$ and $A_{\mu}$, a perturbative process. Defects provide an alternative mechanism. If during the symmetry breaking phase transition defects are formed, then a substantial fraction of the energy is trapped in these defects in the form of topological field configurations of $X$ and $A_{\mu}$. Upon the decay of the defects, the energy is released as $X$ and $A_{\mu}$ quanta which subsequently decay, producing a net baryon asymmetry. Note that defects below the phase transition represent out-of-equilibrium field configurations, thus ensuring that the third Sakharov criterium is satisfied.

As was realized by Kuzmin et al. \cite{KRS}, any net baryon asymmetry produced at very high energies can be erased by baryon number violating nonperturbative processes (sphaleron transitions \cite{sph}) which are unsuppressed above the electroweak scale. Hence, a lot of attention turned to electroweak baryogenesis, the attempt to re-generate a nonvanishing $n_B / s$ by means of sphaleron processes below the electroweak scale, when they are out of equilibrium (see e.g. \cite{revs} for recent reviews).

Topological defects may also play a role in electroweak baryogenesis \cite{BDT}. If new physics just above the electroweak scale generates topological defects in the cores of which the electroweak symmetry is unbroken, then these defects can mediate baryogenesis below the electroweak scale. The defects are out-of-equilibrium field configurations. In their cores, the sphaleron transitions may be unsuppressed, and, typically, C and CP violation is enhanced in the defect walls, thus demonstrating that all of the Sakharov criteria are satisfied.

Detailed studies (see e.g. \cite{status}), however, have shown that without introducing new physics (e.g. supersymmetry with Higgs and stop masses carefully chosen to lie within narrow intervals), electroweak baryogenesis is too weak to be able to generate the observed value of $n_B / s \sim 10^{-9}$. This criticism applies in particular to string-mediated electroweak baryogenesis \cite{Cline}. Thus, at the present time the origin of the observed baryon to entropy ratio remains a mystery.

It is therefore of interest to explore the possibility that the baryon asymmetry may have been generated at the QCD scale via nonperturbative processes, without the need to introduce any new physics beyond the standard model, except for a solution of the strong CP problem. Since baryon number is globally conserved in QCD, the only way to produce a baryon asymmetry is via charge separation. It is to a discussion of such a mechanism to which we now turn.

\section{QCD Domain Walls}

Crucial for our scenario is the existence of QCD domain walls, a consequence of the recent improved understanding of the vacuum structure of QCD.

Let us first consider a $SU(N)$ gauge theory in the absence of fermions. Since the gauge configuration space has nontrivial topology, there are discrete states $|n>$ which minimize the energy in each of the subspaces of field configuration with winding number $n$ (an integer), and from which in turn the $\theta$ vacua $|\theta>$ can be constructed:
\begin{equation} \label{eq1}
|\theta> \, = \, \sum_n e^{i n \theta} |n> \, .
\end{equation}
By construction, physical quantities, in particular the ground state energy, must be $2 \pi$-periodic in $\theta$. The effects of $\theta \neq 0$ can be recast into an additional term in the QCD Lagrangian:
\begin{equation} \label{eq2}
{\cal L}_{QCD} \, = {1 \over {4 g^2}} Tr F_{\mu \nu} F^{\mu \nu} \, + \,
{{\theta} \over {16 \pi^2}} \epsilon^{\mu \nu \alpha \beta} Tr F_{\mu \nu} F_{\alpha \beta} \, ,
\end{equation}
where $F_{\mu \nu}$ is the gauge field strength tensor and $g$ is the coupling constant.

In the large N limit, the combination $\lambda = g^2 N$ must be held constant. Hence, in this limit the vacuum energy as a function of N and $\theta$ must scale as
\begin{equation} \label{eq3}
E(\theta, N) \, = \, N^2 h(\theta/N) \,
\end{equation}
where $h$ is a continuous function. It is also known \cite{VVWa,VVW} that $E(\theta, N)$ is nontrivial (and proportional to $\theta^2$) for small values of $\theta$.
It is difficult to reconcile this with (\ref{eq3}) and with the $2 \pi$-periodicity of $E(\theta)$ unless $E(\theta)$ has a multi-branch structure
\begin{equation} \label{eq4}
E(\theta) \, = \, N^2 min_k h((\theta + 2 \pi k)/N) \, ,
\end{equation}
where $h$ is a smooth function. Such a multi-branch structure was first proposed for supersymmetric QCD \cite{KS,Witten2}. In a functional integral approach, the prescription corresponds to summation over all branches in a multi-valued effective Lagrangian. In the thermodynamic limit, only the branch with the lowest energy contributes.

Let us now include fermions. In the low energy limit, only the pions $\pi^a$ and the $\eta^{\prime}$ meson contribute. They can be described in terms of the matrix
\begin{equation} \label{eq5}
U \, = \, exp\bigl[ i \sqrt{2} {{\pi^2 \lambda_a} \over {f_{\pi}}} + i {2 \over {\sqrt{3}}} {{\eta^{\prime}} \over {f_{\eta^{\prime}}}} \bigr] \, ,
\end{equation}
where $\lambda_a$ are the Gell-Mann matrices and $f_{\pi}$ ($f_{\eta^{\prime}}$) is the pion ($\eta^{\prime}$) coupling constant.

According to the anomalous Ward identities, for massless quarks ($m_q = 0$) the ground state energy as a function of $\theta$ and $U$ can only depend on the combination 
$\theta - i Tr log U$
\begin{equation} \label{eq6}
E(\theta, U) \, = \, E(\theta - i Tr log U) \, .
\end{equation}
{F}rom (\ref{eq6}) we can immediately derive the form of the effective potential for $U$ for a fixed value of $\theta$:
\begin{equation} \label{eq7}
V_{eff}(U) \, = \, E(\theta, U) \, = \, E(\theta - i Tr log U) \, ,
\end{equation}
which inherits the multi-branch structure of the function $E(\theta)$ of the pure gauge theory. In particular, there are distinct ground states. In the chiral limit, their energies are degenerate, but for finite quark masses the degeneracy is softly broken. The energy density barrier between neighboring vacua is of the order $\Lambda_{QCD}^4$ which is much larger than the energy density difference between the minima which is \cite{FHZ} $\delta \rho \sim m_q \Lambda_{QCD}^3$ ($\Lambda_{QCD}$ is the QCD symmetry breaking scale, about $200$MeV). The walls are described by a tension $\sigma$ which is of the order $\Lambda_{QCD}^3$.

The presence of degenerate minima of $V_{eff}(U)$ leads to the existence of domain walls separating regions in space where $U$ has relaxed into different minima. Such domain walls will inevitably be formed \cite{Kibble} during the QCD phase transition.

As first realized in \cite{HZ}, the presence of QCD domain walls implies that the second and third Sakharov criteria are automatically satisfied. Note that perturbative QCD processes in the different minima are the same modulo the value of the strong CP parameter which is shifted by $\Delta \theta_j = i Tr log U_j$ in the j-th minimum $U_j$. In order to avoid the strong CP problem, the effective value $\theta_{eff}$ must be close to zero in the global minimum at the present time. In this case, $\theta_{eff}$ will be of the order 1 in the meta-stable vacua. Hence, there is (almost) maximal CP violation across the domain walls. Note that no new physics is required in order to generate a large amount of CP violation. As stressed in the Introduction, the domain walls are out-of-equilibrium configurations. Hence, Sakharov's second and third criteria are satisfied.

\section{Induced Charge on the Domain Wall}

It has been known for a long time \cite{JR} that solitons can acquire fractional fermionic charges. The prototypical example is a $1+1$-dimensional theory with fermions coupling to a real scalar field with a double well potential via a Yukawa coupling term. In the background of the kink solution for the scalar field, the effective Lagrangian for the fermions $\psi$ is
\begin{equation} \label{lag2}
{\cal L}_2 \, = \, \bar{\psi}(i\partial_{j}\tau^{j}
- me^{i\alpha(z)\tau_3})\psi \, ,
\end{equation}
where $\alpha(z)$ parameterizes the kink, $z$ is the spatial coordinate, and $\tau_i$ denote the Pauli matrices. In this example, the induced fermion charge $B^{(2)}$ of the ground state is given by the net change of $\alpha(z)$:
\begin{equation} \label{charge2}
B^{(2)} \, = \, \int\bar{\psi}\gamma_0\psi dz \, = \, {{\Delta \alpha} \over {2 \pi}} \, ,
\end{equation} 
where $\Delta \alpha = \alpha(+\infty) - \alpha(-\infty)$.

In a similar way, domain walls in a $3+1$-dimensional theory can acquire a fermionic charge. Because of the planar symmetry, the computation can be reduced to that of the above $1+1$-dimensional model. The starting point is the following simplified Lagrangian for the nucleon $N$ interacting with the non-fluctuating chiral field $U$:
\begin{equation} \label{lag4}
{\cal L}_4 \, = \, \bar{N} i\partial_{\mu}\gamma^{\mu}N - 
m_{N} \bar{N}_{L} U N_{R} - m_{N} \bar{N}_{R} U^{+} N_{L}  - 
\lambda (\bar{N}_{L} N_R)(\bar{N}_{R} N_L) \, ,
\end{equation} 
where $m_N$ is the nucleon mass. The last term is a four-fermion interaction term, and $\lambda > 0$ corresponds to repulsion in the $U(1)$ channel. The $3+1$-dimensional charge is given by
\begin{equation} \label{charge4}
B^{(4)} \, = \, \int\bar{N}\gamma_0 N d^3x \, .
\end{equation}

In order to reduce the computation of the four-dimensional charge $B^{(4)}$ to the two-dimensional problem of (\ref{lag2}) and (\ref{charge2}), we write the four-dimensional spinors $N_R$ and $N_L$ in terms of a set of two-component spinors. Note that unless $\lambda \neq 0$, the contributions to $B^{(4)}$ from different two-component spinors cancel (details will be given elsewhere \cite{BHZ2}).

{F}or simplicity, we consider a wall separating the true vacuum from its neighboring meta-stable ground state with $- i Tr log U > 0$. In the following section we will discuss under which circumstances these walls dominate over all other possible walls.

Because of the coupling of $\psi$ to $U$, the effective nucleon mass $m_{eff}$ takes on different values in different vacua $k$:
\begin{equation}
m_{eff}(k) \, = \, m_N \, + \, m_q f(\theta, k) \, ,
\end{equation}
where the function $f(\theta, k)$ depends on the precise form of the effective potential $V_{eff}(U)$ \cite{BHZ2}. In the adiabatic approximation, $m_{eff}$ should be considered as a slowly varying function of $z$ (the direction orthogonal to the domain wall). Fortunately, only the asymptotic values of $m_{eff}$ enter the final answer. The resulting expression for the two-dimensional baryon number of the wall is
\begin{equation}
B^{(2)} \, = \, {1 \over {\pi}} arccos {{m_{eff}} \over {\tilde \lambda}} \vert_{z=-\infty}^{z=+\infty} \, ,
\end{equation}
where ${\tilde \lambda}$ has dimension of mass and can be found in terms of $\lambda$ and $m_N$. For the domain wall we are considering, $B^{(2)}$ is negative if we take the false vacuum to be at $z = + \infty$ and the true vacuum at $z = - \infty$. 

To find the original four-dimensional baryon charge $B^{(4)}$, we should take account of the degeneracy
related to the symmetry under shifts  along the wall plane. In many body physics
the definition of the charge is $B=\int\sum_i\bar{N^i}
\gamma_0 N^i d^3x$
where the sum runs over all particles
(possible quantum  states). In our specific case this summation 
leads to the result
\begin{equation} \label{1}
B^{(4)} \, = \, B^{(2)}g\int\frac{dxdydp_xdp_y}{(2\pi)^2}\equiv B^{(2)}N \; ,
\end{equation}
where $g=4$ describes the degeneracy in spin and isospin. In the following, we shall estimate the value of $N$.

For a wall with surface area $S$ and for a fixed number of quantum states $N$ we have 
\begin{equation}
N \, = \, g S\int\frac{d^2p}{(2\pi)^2} \, = \, \frac{g S p_F^2}{4\pi} \, ,
\end{equation}
where $p_F$ is the Fermi momentum.
The Fermi energy $\bar{E}_F$ of the domain-wall fermions is determined  by 
\begin{equation} \label{2}
\bar{E}_F \, = \, gS\int\frac{pd^2p}{(2\pi)^2}=\frac{2}{3}Np_F \, = \,
\frac{4}{3} \, \sqrt{\frac{\pi}{g}}\frac{N^{3/2}}{\sqrt{S}}.
\end{equation}
The total energy of the fermions residing on the surface $S$ is given by 
\begin{equation} \label{3}
\bar{E_0} \, = \, \sigma S+
\frac{4}{3} \, \sqrt{\frac{\pi}{g}}\frac{N^{3/2}}{\sqrt{S}},
\end{equation}
The size of the surface which can accommodate the fixed number 
of fermions $N$ can be found from the minimization equation
\begin{equation}
\frac{d \bar{E}_{0}}{ dS}|_{N=const} \, = \, 0 \, . 
\end{equation} 
which relates the density of fermions per unit area $n=\frac{N}{S}$ to the 
wall tension $\sigma$:
\begin{equation} \label{4}
n^{3/2} \, = \, \sigma\sqrt{\frac{9g}{4\pi}}.
\end{equation}
Hence, the induced fractional
charge on the domain wall follows from Eqs. (\ref{1}) and (\ref{4}):
\begin{equation} \label{5}
Q \, = \, B^{(4)} \, = \, - |B^{(2)}| N \, = \,
-S |B^{(2)}| \sigma^{2/3}(\frac{9g}{4\pi})^{1/3} \; ,
\end{equation}
which can be expressed in terms of a dimensionless
constant $\alpha_1$:
\begin{equation} \label{charge}
Q \, = \, - S\Lambda_{QCD}^2\alpha_1, ~~\alpha_1 \, =  \,
\frac{\sigma^{2/3}}{q\Lambda_{QCD}^2}(\frac{9g}{4\pi})^{1/3}|B^{(2)}|.
\end{equation}

\section{B-Shell Formation and Baryogenesis}

We now turn to our proposed baryogenesis (charge separation) scenario. At the QCD phase transition at the temperature $T_c \simeq \Lambda_{QCD}$ the chiral condensate $U$ forms. Because of the presence of nearly degenerate states, a network of domain walls will arise immediately after $T_c$ by the usual Kibble mechanism \cite{Kibble}. At $T_c$, the energy difference between vacua is negligible compared to the energy in thermal fluctuations, and hence the 
states are equidistributed. The initial wall separation (correlation length $\xi$) depends on the details of the chiral phase transition but is expected to be microscopic. Below $T_c$, the wall network coarsens. We will assume that an infinite wall network will exist until a temperature $T_d$ at which time the energy difference between correlation volumes of the true vacuum and the false vacuum closest in energy becomes thermodynamically important. At this time, the wall network will decay into a number of finite clusters of the false vacuum which we will call B-shells. 

If $\theta = 0$, there are two degenerate meta-stable states $|B>$ and $|C>$ above the true vacuum of lowest energy $|A>$. For simplicity, we ignore meta-stable states of higher energy. A CP transformation exchanges the states $|B>$ and $|C>$. Hence, the baryon charge of a $A - B$ wall will be opposite of that of a $A - C$ wall, and no baryon number will be left behind in the bulk because the number of $A - B$ walls will be the same as the number of $A - C$ walls.

However, if at the temperature $T_c$ the value of $\theta$ is different from $0$, then the situation is very different. This is the case which will be considered below. Thus, we are making the assumption that the strong CP problem is cured by an axion at a temperature below $T_c$. At $T = T_c$, the axion is not yet in its ground state, and thus $\theta(T_c)$ might be of order unity. Note that as long as the initial value $\theta(T_c)$ is the same in the entire observed Universe, the sign of the baryon asymmetry will also be the same. This will occur if the Universe undergoes inflation either after or during the Peccei-Quinn symmetry breaking. In this case there is a splitting of $\theta(T_c) m_q \Lambda_{QCD}^3$ between the energy densities of the states $|B>$ and $|C>$ which translates into a splitting $\Delta M \sim \theta(T_c) M$ between the masses of B-shells of the phases $|B>$ and $|C>$ with negative and positive baryon numbers (here, $M$ stands for the B-shell mass at $\theta = 0$).
We will assume that $\Delta M$ is larger than $T_d$. Since the correlation length $\xi$ grows rapidly after $T_c$, this requirement can be achieved without requiring a large value of $\theta$. In this case, at the temperature $T_d$, only B-shells of one type, of negative baryon number, will remain. For a value $\xi(T_d) \sim 10^6 T_c^{-1}$ (a value which we argue below is reasonable) and assuming spherically symmetric B-shells, the criterium for $\theta(T_c)$ becomes:
\begin{equation}
\theta(T_c) \, \gg \, \bigl( \xi(T_d) \Lambda_{QCD} \bigr)^{-3} {{T_d} \over {m_q}} \, \sim \, 10^{-16} \, ,
\end{equation}
where in the last step we have replaced $T_d$ by $T_c$ to obtain a conservative bound. 

Note that the typical wall separation $\xi(T_d)$ is rather uncertain since it depends on the initial correlation length at formation, on the details of the damping mechanism and on the interplay between the energy bias and the surface tension in the walls \cite{Kibble,Zurek,Gelmini}. Given the typical wall separation $\xi$, the total area $S$ in walls at the temperature $T_d$ within some reference volume $V$ is $S \sim V / \xi$. The total baryon charge is given by (\ref{charge}). Since the entropy density is $s = g_* T_d^3$, where $g_* \sim 10$ is the number of spin degrees of freedom in the radiation bath at $T_d$, the net baryon to entropy ratio at $T_d$ becomes
\begin{equation} \label{result1}
{{n_B} \over s}(T_d) \, \sim \, {{\alpha_1 \Lambda_{QCD}^2} \over
{g_* \xi(T_d) T_d^3}} \, .
\end{equation}

The evolution of the B-shells after $T_d$ requires a detailed study. Qualitatively, we expect that the bubbles will shrink, but not decay completely since they will eventually be stabilized by the fermions. We expect the annihilation cross-section between a baryon and a B-shell to be suppressed by a large power of the ratio of the Compton wavelength of the baryon and the radius of the B-shell. As the non-relativistic baryons can hardly cross the wall, we expect the shells to be stable against the escape of baryons from the interior, but able to lose heat by baryon pair annihilation and emission of the photons and/or neutrinos. The quantum stability of the B-shells will be addressed separately \cite{BHZ2}. Generally, one expects an exponential suppression of quantum decays by the baryon charge of the surface.

To proceed, we introduce two dimensionless constants $\alpha_2$ and $\alpha_3$, parameterizing the total area and volume of the B-shells as $\alpha_2^2 V / \xi$ and $\alpha_3 V$, respectively. This parameterization does not imply any specific assumption about the form of the B-shells. Neglecting the expansion of the Universe between $T_c$ and $T_d$ we obtain 
\begin{equation} \label{result2}
{{n_B} \over s}(T_d) \, \sim \, {{\alpha_1 \alpha_2^2 \Lambda_{QCD}^2} \over
{g_* \xi(T_d) T_d^3}} \, .
\end{equation}
Since the energy density $\rho_B$ in B-shells will red-shift as matter, the   contribution $\Omega_B$ of B-shells to the dark matter of the Universe is given by
\begin{equation}
\Omega_B \, \simeq \, {{\rho_B(t_{eq})} \over {\rho_r(t_{eq})}} \, =  ,\ {{\rho_B(t_{eq})} \over {g_* T_d^3 T_{eq}}} \, ,
\end{equation}
where $t_{eq}$ is the time of equal matter and radiation. Assuming that the B-shell energy is dominated by the false vacuum energy, we obtain:
\begin{equation} \label{result3}
\Omega_B \, \sim \, \alpha_3 {{m_q T_c^3} \over {T_{eq} T_d^3}} \, .
\end{equation}
Comparing (\ref{result2}) and (\ref{result3}), we see that the resulting values of $n_B / s$ and of $\Omega_B$ are related with each other via the geometric parameters $\alpha_2$ and $\alpha_3$. At the moment, we are not able to calculate these parameters directly. However, we can reverse the argument and ask what values of $\alpha_2$ and $\alpha_3$ are required in order to explain both $\Omega_B \sim 1$ and $n_B / s \sim 10^{-10}$. Taking $T_d \sim T_c$ we obtain $\alpha_3 \sim 10^{-6}$ and $\alpha^2 \sim 10^{-6} \xi T_d$. Since by definition $\alpha_2$ and since $\xi T_d > 1$, we are left with the window
\begin{equation} \label{range}
T_c^{-1} \, < \, \xi \, < \, 10^6 T_c^{-1}  
\end{equation}
for the proposed mechanism to be operative. This window is consistent with the Kibble-Zurek scenario \cite{Kibble,Zurek} of defect formation in the early Universe.

\section{Discussion}

We have seen that it is possible, without fine tuning of parameters, to obtain a reasonable value of the baryon to entropy ratio in the bulk. B-shells will contribute to the dark matter of the Universe, and there is a region of parameter space for which B-shells will make up the bulk of the dark matter.
Note that in our charge separation scenario, charges are separated only over microscopic scales.

For the scenario to work, it is important that the B-shells be stable. We  argue that Fermi pressure will stabilize the shells against collapse. We assume that eventually the radius will be much larger than $\Lambda_{QCD}^{-1}$, thus justifying the use of the results for the charge per unit area derived for a flat surface. In this case, the total  energy for a spherical shell of radius $R$ is  
\begin{equation} \label{energy}
E \, = \, 4 \pi \sigma R^2 + {4 \over 3} \sqrt{{{\pi} \over g}} {{N^{3/2}} \over {2 \sqrt{\pi} R}}
+ {{4 \pi R^3} \over 3} \delta \rho 
\end{equation}
(see (\ref{3})). Here, $N$ is the total baryon number of the shell which is held fixed and can be estimated from the initial shell radius $R_0 = \xi$, 
$N  = \, 4 \pi \xi^2 n$,
where the baryon number density $n$ is given by (\ref{4}). By minimizing (\ref{energy}) with respect to $R$ (and noting that the surface tension term is negligible compared to the other two terms), we can find the stabilization radius for a fixed initial baryon charge
\begin{equation}
R^4 \, \simeq \, {{N^{3/2}} \over {6 \pi \sqrt{g} \Delta \rho}} \, ,
\end{equation}
which then also determines the total energy of a B-shell, with the result that $E \sim N^{9/8}$. Taking the value of $\xi$ to be at the upper end of the range (\ref{range}) we find $E \sim 10^{14}$GeV or about $10^{-10}$g.

Since the negative baryon charge is trapped in a topological configuration, we expect the annihilation cross section between nucleons and B-shells to be greatly suppressed because of the mismatch between the Compton wavelength of the nucleons and the B-shell radius. Similarly, any charge loss mechanism will also experience this phase space suppression. Constraints on our proposed mechanism based on elastic scattering of B-shells in dark matter detectors remain to be explored. 

In conclusion, qualitative as our arguments are, they suggest that 
baryogenesis can proceed at the QCD scale, and might be tightly
connected with the origin of the dark matter in the Universe.

\section*{Acknowledgments}

One of us (R.B.) would like to thank the organizers of SEWM-98 for their hospitality. We are grateful to P. Arnold, E. Mottola, M. Shaposhnikov, E. Shuryak, A. Vilenkin and L. Yaffe for valuable comments. This work was supported in part by the Canadian NSERC and by the U.S. Department of Energy under 
Contract DE-FG02-91ER40688, TASK A.

\end{document}